# Luminescent emission of excited Rydberg excitons from monolayer $WSe_2$

*Shao-Yu Chen,[1] Zhengguang Lu,[2,3] Thomas Goldstein,[1] Jiayue Tong,[1] Andrey Chaves[4], Jens Kunstmann[5], L. S. R. Cavalcante[4], Tomasz Woźniak[5,6], Gotthard Seifert[5], D. R. Reichman,[7] Takashi Taniguchi,[8] Kenji Watanabe,[8] Dmitry Smirnov[2] and Jun Yan[1,*]*

[1]Department of Physics, University of Massachusetts, Amherst, Massachusetts 01003, USA

[2]National High Magnetic Field Laboratory, Tallahassee, Florida 32310, USA

[3]Department of Physics, Florida State University, Tallahassee, Florida, 32306, USA

[4] Departamento de Física, Universidade Federal do Ceará, Caixa Postal 6030, Campus do Pici, 60455-900 Fortaleza, Ceará, Brazil

[5] Theoretical Chemistry, Department of Chemistry and Food Chemistry & Center for Advancing Electronics Dresden, TU Dresden, 01062 Dresden, Germany

[6] Department of Theoretical Physics, Wrocław University of Science and Technology, wyb. Wyspiańskiego 27, 50-370, Wrocław, Poland

[7] Department of Chemistry, Columbia University, New York, New York 10027, USA

[8] National Institute of Materials Science, 1-1 Namiki, Tsukuba, Ibaraki 305-0044, Japan

[*]Corresponding Author: Jun Yan. Tel: (413)545-0853 Fax: (413)545-1691

E-mail: yan@physics.umass.edu

**Abstract**

We report the experimental observation of radiative recombination from Rydberg excitons in a two-dimensional semiconductor, monolayer $WSe_2$, encapsulated in hexagonal boron nitride. Excitonic emission up to the $4s$ excited state is directly observed in photoluminescence spectroscopy in an out-of-plane magnetic field up to 31 Tesla. We confirm the progressively larger exciton size for higher energy excited states through diamagnetic shift measurements. This also enables us to estimate the $1s$ exciton binding energy to be about 170 meV, which is significantly smaller than most previous reports. The Zeeman shift of the $1s$ to $3s$ states, from both luminescence and absorption measurements, exhibits a monotonic increase of $g$-factor, reflecting nontrivial magnetic-dipole-moment differences between ground and excited exciton states. This systematic evolution of magnetic dipole moments is theoretically explained from the spreading of the Rydberg states in momentum space.

Keywords: Rydberg exciton, tungsten diselenide, diamagnetic shift, Zeeman shift, magnetic dipole moment.

Exciton physics in layered hexagonal transition metal dichalcogenide (H-TMDC) semiconductors has attracted much interest in recent years. In its monolayer (1L) form, the H-TMDC atomic layers possess a direct bandgap, despite the fact that the bulk crystals are indirect gap semiconductors[1,2]. These direct gaps are located at two inequivalent K valleys of the Brillouin zone, providing facile access to the valley degree of freedom [3]. Under light illumination, optically excited electrons and holes relax to the band edge of the valleys and form various bound states. The ground-state 1*s* bright exciton, a charge-neutral spin-zero composite boson with near-zero center-of-mass momentum, can be viewed as a benchmark optical feature of 1L-TMDC [1,2] . With a size and binding energy that lies in-between what is found for Wannier-Mott and Frenkel-type excitons, these electron-hole excitations have prominent manifestations in both emission and absorption even at room temperature. For energies lying below the 1*s* bright exciton, stable bound states composed of three-, four-, and five-particles can form and have been identified experimentally[4–8]. Above the 1*s* exciton and below the quasi-particle bandgap, a clear energy window of a few hundred meV opens and allows for studies of excited Rydberg states confined in the 2D atomic layers.

Rydberg states are intriguing entities that carry quantized angular momentum, and inherit the valley degree of freedom from the TMDC electronic bands [9]. By experimental techniques such as differential reflectance, photoluminescence excitation, and two-photon absorption, excited two-particle states with *s* and *p* symmetries have been identified[10–16]. Meanwhile, these identifications have not always been consistent, and it is desirable to have more accurate experimental probes to achieve an extensive understanding of excitonic Rydberg states in 2D materials[17].

In this Letter, we study excited Rydberg excitons in monolayer tungsten diselenide (1L-$WSe_2$) using photoluminescence (PL) spectroscopy. Despite being a highly accurate and popular technique, PL spectroscopy typically focuses on energies at and below the 1*s* bright exciton, and is infrequently used (compared to say, reflection spectroscopy) for investigating excited Rydberg excitons in 2D-TMDCs due to Kasha's rule[18]. Specifically, the excited Rydberg excitons at higher energy have more decay channels than the ground state. Once formed inside the light cone, the radiative recombination of excited Rydberg excitons faces competition from scattering by disorder, charges, phonons and other excitons, which can send them outside the light cone or transition them to lower energy states. With continuous improvement of sample quality that suppresses nonradiative decay channels, several studies have successfully revealed the 2*s* exciton PL emission at low temperatures in 1L-$WSe_2$ recently [18–20].

Here, with the aid of a strong external magnetic field up to 31 Tesla and a high-quality sample encapsulated in hexagonal boron nitride (hBN), we successfully resolve the magneto-PL of 1*s*, 2*s*, 3*s* and 4*s* excitons. This allows us to determine their Zeeman and diamagnetic shifts with high accuracy. Combined with theoretical calculation by assuming the Rytova-Keldysh electron-hole interaction potential [21,22], we infer a 1*s* binding energy of about 170 meV. Previous studies with different experimental techniques have found varied binding energies ranging ~0.2–0.8 eV[11,23–26]. Our smaller value is likely due, at least in part, to additional dielectric screening of the excitons provided by the encapsulation of our sample in hBN, implying the high sensitivity of the dielectric screening effect on Coulomb interaction in 2D material system.

We further verify our PL results *via* the use of reflection spectroscopy and find that the high accuracy of our measurements enables us to minimize uncertainties in determining the Zeeman shift of Rydberg excitons. Due to the different sizes of the 1*s*, 2*s*, 3*s* and 4*s* excitons, which are reflected in the drastically distinct diamagnetic shifts that we measure, the spreading of the exciton envelope wavefunctions in momentum (k) space is smaller for Rydberg excitons with higher principal quantum number. Compared to, e.g., 2*s* and 3*s*, the 1*s* exciton thus samples optical transitions from electronic states covering a larger area near the +K and –K points in momentum space. As a result, the $g$-factors of the 1*s*, 2*s* and 3*s* excitons are observed to increase progressively. We calculate the magnetic dipole moment in momentum space and find that the two contributions due to atomic orbitals and self-rotation of the Bloch wave functions indeed decrease away from +K and –K, in agreement with our experimental observations.

The high-quality hBN-sandwiched 1L-$WSe_2$ sample used in our experiment is made by a dry transfer technique; see Fig.1a for an optical microscope image. We mount the sample on a fiber-based custom microscope system equipped with 3-axis piezo stages as illustrated in Fig.1b. The whole optical setup is cooled down to 2 K in a cryostat integrated with a 31 T resistive magnet, where the magnetic field is applied perpendicular to the atomic layer. The sample is excited by 2.33 eV unpolarized laser light in back-scattering geometry to achieve equal optical excitation in the +K and –K valleys. In the collection path, we employ a thin-film broadband quarter waveplate ($\lambda/4$) and a linear polarizer (LP) to selectively collect the PL signals with $\sigma^-$ helicity which originates from optical emission from the –K valley. Finally, the collected light is passed through a long pass filter (LPF), coupled to a spectrometer through another fiber and detected by a thermoelectrically-cooled CCD camera.

Assisted by the passivation from hBN flakes, our sample is of high quality and has only minor electron doping (see Supporting Information). This enables us to observe the charge neutral Rydberg excitons throughout the measurements. Figure 1c show the typical PL spectra of our sample at ±31 Tesla. Around 1.73 eV we observe the 1$s$ bright exciton emission that dominates the spectrum. The $\sigma^-$ 1$s$ PL has a higher energy at positive compared to negative magnetic fields, indicating that the Zeeman shift is positive for –K bright excitons, consistent with conventions established in most of previous studies[27–30]. At higher energy, between 1.85 and 1.95 eV, we observe three additional peaks with similar positive Zeeman shifts. These peaks have decreasing intensity at higher energy and are attributed to the 2$s$, 3$s$ and 4$s$ excitons, respectively. The full width at half maximum (FWHM) of the peaks are: 1$s$, 5.7 meV; 2$s$, 6.2 meV; 3$s$, 7.4 meV and 4$s$, 9.0 meV at –31T, consistent with the fact that the Rydberg states with higher quantum number have more decay channels and thus a shorter lifetime. Quantitatively, our relatively strong 2$s$ PL, combined with the its smaller oscillator strength (~1/12 that of 1$s$), suggests that the 2$s$ population decays 5 times faster than 1$s$. We note that at 31 T, some $\sigma^+$ signal at lower energy leaks out because of the imperfect magnetic response of the quarter waveplates in the collection path at high magnetic fields over a large energy range. The $\sigma^+$ PL arises from emissions in the +K valley and its energy at 31 T matches that of the $\sigma^-$ –K emission at –31 T, reflecting that the +K and –K excitons are time reversal pairs with opposite Zeeman shifts.

Figure 1d displays the 2D map of PL intensity as a function of magnetic field $B$ ranging from –31 to 31 T. For the 1$s$ exciton peak shown in the left subpanel at about 1.73 eV, the magnetic response is dominated by the linear Zeeman shift. For 2$s$, 3$s$ and 4$s$ states between 1.85 and 1.95 eV however, the peak evolution becomes increasingly curved, indicating that an additional contribution from the quadratic diamagnetic shift becomes more important. At our maximum field of 31 Tesla, the cyclotron energy $\hbar\omega_c \approx \hbar eB/m_e$ of the electrons is about 7.2 meV, which is much smaller than the binding energy of the exciton as we extract below. The magnetic field dependence of the exciton energy can thus be described by a summation of the Zeeman shift and the diamagnetic shift as

$$E(B) = E_0 - g\mu_{\mathrm{B}}B + \frac{e^2}{8m_{\mathrm{r}}}\langle r^2\rangle B^2, \qquad (1)$$

where $E_0$ is the exciton energy at zero field, $\mu_B$ is the Bohr magneton, $m_r = \frac{m_e m_h}{m_e + m_h}$ is the reduced mass of the exciton, and $\langle r^2 \rangle = \langle \Psi | r^2 | \Psi \rangle$ is the expectation value calculated over the exciton's wavefunction, which provides a measurement of the exciton size. Note that for an exciton with magnetic dipole moment $\vec{\mu}$ in an out-of-plane magnetic field $\vec{B}$, the Zeeman shift is given by $E_z = -\vec{\mu} \cdot \vec{B} = -g \mu_B B$. This indicates that the $g$-factor of –K exciton is negative. In literature [31–35], the 1$s$ exciton $g$-factor has alternatively been defined as the energy difference of +K and –K excitons normalized by $\mu_B$. Numerically it is equal to twice the value of our –K valley exciton $g$-factor defined here.

Assuming that $m_r$ does not vary significantly, the curvatures seen in Fig.1d are consistent with the fact that Rydberg excitons with larger quantum numbers have larger sizes. 1L-$WSe_2$, when sandwiched between hBN, exhibits rich exciton-phonon interaction effects [36,37]. Optical features at energies around our 2$s$ exciton have been previously interpreted as a $WSe_2$-hBN phonon replica of the 1$s$ exciton. Here our observed distinct diamagnetic shifts provide firm evidence that the PL emission at ~1.86 eV is indeed from the 2$s$ exciton as we have previously surmised [18].

2D TMDC excitons have been studied in high magnetic fields before, using both differential reflectance (DR, $\Delta R/R$) and magneto-PL spectroscopy[31–35]. These two techniques have different advantages and disadvantages. In reflectance/absorption spectroscopy, by performing high-order derivatives on heavily averaged and smoothed spectra, seemingly small and subtle features can be made visible, and excited Rydberg states have been revealed this way even at zero magnetic field and room temperature[10,11]. However, due to the multi-layer structures (e.g. $SiO_2$ and hBNs) that cause multiple reflections and interference, the spectra are typically asymmetrically distorted and have large sloping background, making accurate determination of Rydberg state energies challenging. PL typically gives much better-defined emission peaks, rendering peak position assignments more straightforward. However, due to Kasha's rule, radiative emission from excited excitons is difficult to observe, as a result of their lower population density of excitons and their smaller radiation dipole moment, as well as the strong competition from other intrinsic and extrinsic decay channels. Another disadvantage of PL spectroscopy is the typical existence of a Stokes shift between PL and absorption due to disorder [38]. This is especially important in the context of our investigation of Zeeman and diamagnetic shifts, as the unknown magnetic field dependent Stokes shift may complicate the interpretation of data. These disadvantages of PL spectroscopy are significantly alleviated in our device, as discussed in detail below, due to its superior sample quality. This is evidenced in part by the appearance of luminescence of higher Rydberg states and a narrow FWHM, indicating minimal extrinsic scattering.

To ensure the reliability of our measurement and analysis, we performed a control study using a 17 T superconducting magnet integrated with an optical cryostat that allows for free space light coupling, enabling us to compare PL and DR measurements on the same sample. In Figure 2a, we plot the PL spectra along with the DR spectra as well as its 2$^{nd}$ derivative (2DDR) at 5 K and 17 T. The 1$s$ exciton has strong signal in both PL and reflectance, and its energy as determined by PL, DR and 2DDR are highly consistent with each other, manifesting negligible Stokes shift. This indicates that our PL spectroscopy is as good as reflectance in determining the 1$s$ exciton energy, as well as the Zeeman and Stokes shifts.

For excited Rydberg states, we observe 2$s$ and 3$s$ exciton peaks clearly in the PL spectrum. The differential reflectance spectrum also resolves well the 2$s$ and 3$s$ absorption dips. However, there is a large sloping background and a significant distortion due to the interference effects induced by the multiple dielectric layers in our hBN/1L-$WSe_2$/hBN sample on an oxidized silicon chip, rendering assignment of the absolute peak positions less accurate. The sloping background can be removed by performing the 2$^{nd}$ derivative of the $\Delta R/R$ spectra, and we can extract the peak energy by fitting the dominant peak with Gaussian functions; see Fig. 2a. However, the asymmetry in the spectrum makes determination of the absolute peak energy less reliable, which causes an artificial blueshift compared to PL. Further in the energy range of the 3$s$ state, several small, albeit sharp artifacts show up in the 2DDR spectra, making the accurate extraction of the 3$s$ dip position challenging. We thus conclude that PL is more accurate in determining 2$s$ and 3$s$ energies at high magnetic fields. At low magnetic fields in the range of ±5 T, we found that the PL spectral weight above 2$s$ cannot be fully attributed to 3$s$ and 4$s$ excitons, in contrast to previous studies[19]. This can be seen in the 2D map in Fig.1d: by extending the 3$s$ exciton position from high $B$ to low $B$ (black dashed curve), it is clear that there is some additional spectral weight between 3$s$ and 2$s$ that disappears at high magnetic fields, the origin of which is currently unclear.

Quantitatively, according to Eq.(1), we can find the –K exciton Zeeman shift by calculating $E_z(B) = -g\mu_{\mathrm{B}}B = \frac{E(B)-E(-B)}{2}$, and the diamagnetic shift by $E_{\mathrm{avg}}(B) = E_0 + E_{\mathrm{dia}}(B) = \frac{E(B)+E(-B)}{2}$. In Figs.2b and 2c, we compare the values of the 2$s$ $E_{\mathrm{Z}}(B)$ and $E_{\mathrm{avg}}(B)$, respectively, as determined from PL and 2DDR. The experimental results from the two different methods are highly consistent except for the ~1meV difference in $E_0$ as discussed above. In Fig.2b, the Zeeman shift values from the two types of measurement overlap with each other, giving the same slope with an uncertainty of less than 2%. In Fig.2c, both data sets can be well fit by quadratic curves with the same curvature with an uncertainty of about ~ 4% in the coefficient (in fact, the two fits in Fig.2c are made with the same quadratic coefficient).

Given that our sample has a negligible Stokes shift from the above analysis, we focus below mostly on the magneto-PL data which are measured up to 31 Tesla. In Fig.3a, we plot $E_{\mathrm{avg}}$ of the $\sigma^-$ PL as a function of $B^2$. Defining $E_{\mathrm{dia}} = \alpha B^2 = \frac{e^2}{8m_{\mathrm{r}}}\langle r^2\rangle B^2$, we note that $\mathrm{E}_{\mathrm{avg}}(B) = E_0 + \alpha B^2$. The slope of our data gives $\alpha$ and the $B = 0$ T intercept gives $E_0$. We find $\alpha$ to be 0.5, 5.8, and 17.6 $\mu eV/T^2$, and $E_0$ to be 1.727(1), 1.858(1), and 1.884(1) eV for the $1s$ to $3s$ exciton states, respectively.

The values of $\alpha$ and the reduced mass $m_{\mathrm{r}}$ determine the size of the Rydberg excitons $\sqrt{\langle r^2\rangle}$. The mass of electrons and holes have been measured by several different methods in literature. By fitting the cyclotron frequency extracted from Shubnikov-de Haas oscillations in magneto-transport measurements, the effective mass of holes has been estimated as $m_h$ = 0.45 to $0.5m_0$ [39]. Single electron transistor spectroscopy has found both electron and hole masses to be about 0.4–$0.6m_0$ [40]. In a magneto-optical measurement study of the inter-Landau level transition [41], the exciton $m_r$ was estimated to be around 0.27 to $0.31m_0$. These results are consistent with the value reported in *ab initio* calculations [42,43]. In Fig.3b we plot $\sqrt{\langle r^2\rangle}$ as a function of the reduced mass for our measured $\alpha$ values. Assuming a reduced mass $m_{\mathrm{r}} = 0.22m_0$, close to the lower bound expected from the above-cited literature, we can determine the radii of 2.2, 7.6, and 13.3 nm for $1s$, $2s$, and $3s$ excitons, respectively.

The $E_0$ value provides a facile means for estimating the $1s$ exciton binding energy, which has been debated in recent years[11,23–25]. Our $2s$–$1s$ and $3s$–$1s$ energy separations are 131 and 157 meV, respectively. This suggests that the $1s$ exciton binding energy is not much larger than 157 meV. Meanwhile the $2s$–$1s$ separation is less than 8 times the $3s$–$2s$ separation, deviating from the 2D hydrogen model.

The Rydberg exciton in a vector potential created by an external magnetic field can be described by

$$H = \sum_{i=e,h} \frac{(\vec{p}-q_i\vec{A})^2}{2m_i} + V_{eh}(|\vec{r}_e - \vec{r}_h|) \quad (2).$$

Following previous works [10,21,22,44], we have modeled the electron-hole interaction with a Rytova-Keldysh potential

$$V_{eh}(r) = -\frac{e^2}{4\epsilon\rho_0}\left[H_0\left(\frac{r}{\rho_0}\right) - Y_0\left(\frac{r}{\rho_0}\right)\right] \quad (3),$$

where $\epsilon = 4.5\epsilon_0$ is the (static) dielectric constant of hBN, $\rho_0 = 2\pi\frac{\chi_{2D}}{\epsilon}$ with $\chi_{2D} = 7.18$ Å, and $H_0$ and $Y_0$ are the Struve and Neumann functions, respectively. Assuming a reduced mass of 0.22 $m_0$ and a quasi-particle bandgap of 1.9 eV, we numerically calculate the Rydberg exciton energy as a function of external magnetic field. The calculated exciton energy varies quadratically as a function of magnetic field, from which we extract the value of $\alpha$ ($\sqrt{\langle r^2\rangle}$) to be 0.25 (1.6), 4.18 (6.5) and 21.6 (14.7) $\mu eV/T^2$ (nm) respectively for 1*s*, 2*s* and 3*s* states. These values are in reasonable agreement with our experimental results (upper triangles in Fig.3b). An independent calculation found 1*s* exciton $\alpha$ to be 0.08 $\mu eV/T^2$ on a $SiO_2$ substrate[45]. The exciton energies are found to be 1.731 eV (1*s*), 1.859 eV (2*s*), and 1.882 eV (3*s*), in excellent agreement with experimental data. This also suggests that the binding energy of ground state exciton can be estimated around 170 meV. This is a relatively small value compared to existing literature[11,23–25]. We note that the diamagnetic shifts we observe lend credibility to our Rydberg series assignment, providing multiple check points for modeling the binding energy. A recent DR measurement at high magnetic fields used similar techniques and the binding energy agrees well with our results[34]. Our value is significantly smaller than previous zero field DR results (370 meV) that also make use of the energy of Rydberg series[11]. Nevertheless, we point out that the previous zero-field measurements were performed on samples deposited on a silicon substrate and without hBN encapsulation, which is a more weakly-screening dielectric environment. A theoretical study show that the binding energy in this case is expected to be 295 meV [46]. Thus, this previous study is not in conflict with our results. We note that similarly the Bohr radius also depends on dielectric environment, and can be different on different substrates, as well as in different cavities and wave-guides[47,48].

We now discuss the Zeeman shift of the different Rydberg excitons, which are plotted in Fig.4a with PL data up to 31 T and 2DDR data up to 17 T. As expected, the energy shift is linearly proportional to the magnetic field. Interestingly, we observe that the *magnitude* of $g$-factor monotonically increases from 2.15, for the 1$s$ exciton, to 2.53, for the 3$s$ exciton. This systematic increase of $g$-factor for larger excitons is real and is observed for both PL and DR. Our data thus indicate nontrivial differences between the magnetic dipole moments of different Rydberg states.

As has been discussed in several previous studies[27–30], the Zeeman shift of the exciton, $-g\mu_{\mathrm{B}}B$, can be understood as the difference between the Zeeman shift of the conduction band $-\mu_{\mathrm{c}}B$ and that of the valence band $-\mu_{\mathrm{v}}B$. The magnetic dipole moment of the electronic states at K is comprised of three components, originating from a spin term $\mu_s$, an atomic orbital term $\mu_o$, and a term related to the self-rotation of a Bloch wave packet around its center of mass $\mu_{sr}$ [49,50], also known as the inter-cellular or valley contribution [27–30]. Defining $\Delta\mu_n$, where $n = s$, $o$, or $sr$, as the difference of these contributions from the conduction and the valence bands, the exciton $g$-factor is expressed as $g = \frac{1}{\mu_{\mathrm{B}}}\Delta\mu_{total} = \frac{1}{\mu_{\mathrm{B}}}(\Delta\mu_s + \Delta\mu_o + \Delta\mu_{sr})$

For the bright K-valley excitons, the spin of the electron states in the conduction and valence bands point in the same direction and, therefore, their spin contributions cancel out, yielding $\Delta\mu_s = 0$. The atomic orbital contribution is related to the orbital composition of the electron and hole states in the vicinity of K. It is known that these states are mainly formed from the tungsten 3$d$ orbitals [42], where the electrons exhibit approximately zero angular momentum, while for the holes, the dipole moment at +K is approximately $-2\mu_{\mathrm{B}}$. We note that our Density Functional Theory (DFT) calculations produces slightly different results, due to contributions from other atomic orbitals in these states; exactly at +K, we obtain $\mu_o$ to be $-0.056\mu_{\mathrm{B}}$ and $-1.483\mu_{\mathrm{B}}$ for conduction and valence bands respectively, so that, at the band edge, $\Delta\mu_o = 1.427\mu_{\mathrm{B}}$. Notice, however, that in the vicinity of K, the orbital composition of the band states changes, and this value is thus reduced as the Bloch wave vector departs from the band edge, as shown by the blue curve in Fig. 4b.

The self-rotation contribution for valence and conduction bands are calculated via the sum of virtual band transitions as in Ref. [50], where the approximate third-nearest-neighbors three-band tight-binding (TB) Hamiltonian for $WSe_2$ [42] is used for convenience. Notice that for a parabolic band with an effective mass $m_*$, the self-rotation contribution is given by $\mu_{sr} = \frac{m_0}{m_*}\mu_{\mathrm{B}}$. Since the two-band Dirac Hamiltonian of $WSe_2$ leads to electrons and holes with the same effective mass, the self-rotation contribution has previously been approximated to zero, e.g. in Ref. [27]. The precision of our PL results, however, allows for the detection of corrections to this approximation, which calls for a theoretical model where band curvatures (beyond the parabolic approximation) and effective masses are more accurately described. Despite its simplicity, the three-band TB Hamiltonian captures the essential needed physical features. Results are shown as the orange curve in Fig. 4b. The actual values $\mu_{sr}$ for each band exactly at the band edge +K are $4.702\mu_{\mathrm{B}}$ and $3.492\mu_{\mathrm{B}}$, so that $\Delta\mu_{sr}$= $1.210\mu_{\mathrm{B}}$. Note that just like the atomic orbital contribution, the self-rotation contribution also decreases as the wave vector moves away from K.

In principle, these values of the dipole moment suggest that excitonic transitions would eventually exhibit a Zeeman shift with an effective $g$-factor $g(\mathrm{K})$ = 2.637. However, the strongly bound excitons in 2D materials exhibit wave functions with a widely-spread envelope in reciprocal space, so that the exciton samples not only the states precisely at the band edge K, but also in its vicinity, where, as previously shown, the angular momentum is effectively smaller. Taking a Fourier transform of the numerically obtained exciton wave functions, $\Psi_{\mathrm{state}}(k) = \langle k|\Psi_{\mathrm{state}}\rangle$ (state = 1$s$, 2$s$, or 3$s$), the expectation value $g_{\mathrm{state}} = \langle\Psi_{\mathrm{state}}|[\Delta\mu_s + \Delta\mu_o + \Delta\mu_{sr}]|\Psi_{\mathrm{state}}\rangle/\mu_{\mathrm{B}}$ leads to theoretical $g$-factors of $g_{1s}$ = 2.224, $g_{2s}$ = 2.311 and $g_{3s}$ = 2.431, which are quantitatively close to the experimental values observed in Fig. 4a. Notice that excited exciton states are less bound and, therefore, narrower in k-space, as one can verify by their wave functions in Fig. 4c. Therefore, $g$-factors approach the $g(\mathrm{K})$ = 2.637 value as the state index increases. In fact, the $g_{1s}$ = 2.15 value observed here agrees well with that reported for the 1$s$ exciton e.g. in Ref. [28], while the experimental findings reported here extend this result to demonstrate state-dependent $g$-factors for excited exciton states.

In conclusion, we have observed 1$s$, 2$s$, 3$s$ and 4$s$ exciton photoluminescence in high quality hBN-encapsulated monolayer $WSe_2$. The superior sample quality enables us to accurately determine the Zeeman and diamagnetic shifts of different Rydberg excitons. We estimate the 1$s$ exciton binding energy to be about 170 meV in this dielectric environment. Evolution of the Rydberg exciton wavefunctions in both real space and momentum space are found to impact their magnetic response, with the former manifested in the diamagnetic shift, and the latter in the Zeeman shift, i.e. magnetic dipole moment size. These findings provide a deeper understanding of the magnetic properties of excitonic states in atomically thin semiconductor materials.

**Supporting Information Available:** Details of theoretical modeling and sample to sample variation of PL spectra. This material is available free of charge via the Internet at http://pubs.acs.org.

**Notes**

The authors declare no competing interests.

**Acknowledgments**

This work is supported mainly by the University of Massachusetts Amherst, and in part by NSF ECCS-1509599. J.Y. thanks Dr. Xiaoxiao Zhang (Cornell University) for help with sample design. Z.L and D.S. acknowledge support from the US Department of Energy (DE-FG02-07ER46451) for magneto-photoluminescence measurements performed at the National High Magnetic Field Laboratory, which is supported by National Science Foundation through NSF/DMR-1157490, DMR-1644779 and the State of Florida. A.C. and L.S.R.C. acknowledge funding from CAPES and from CNPq, by the PQ and PRONEX/FUNCAP programs. J.K. and G.S. acknowledge funding by the German Research Foundation (DFG) under grant number SE 651/45-1. J.K thanks Tobias Frank (University of Regensburg, Germany) for helpful discussions. D.R.R. acknowledges support from NSF-CHE-1839464. Computational resources were provided by ZIH Dresden under project 'transphemat'. T.W acknowledges funding by Polish Ministry of Science and Higher Education within Diamond Grant D\2015 002645. K.W. and T.T. acknowledge support from the Elemental Strategy Initiative conducted by the MEXT, Japan and the CREST (JPMJCR15F3), JST.

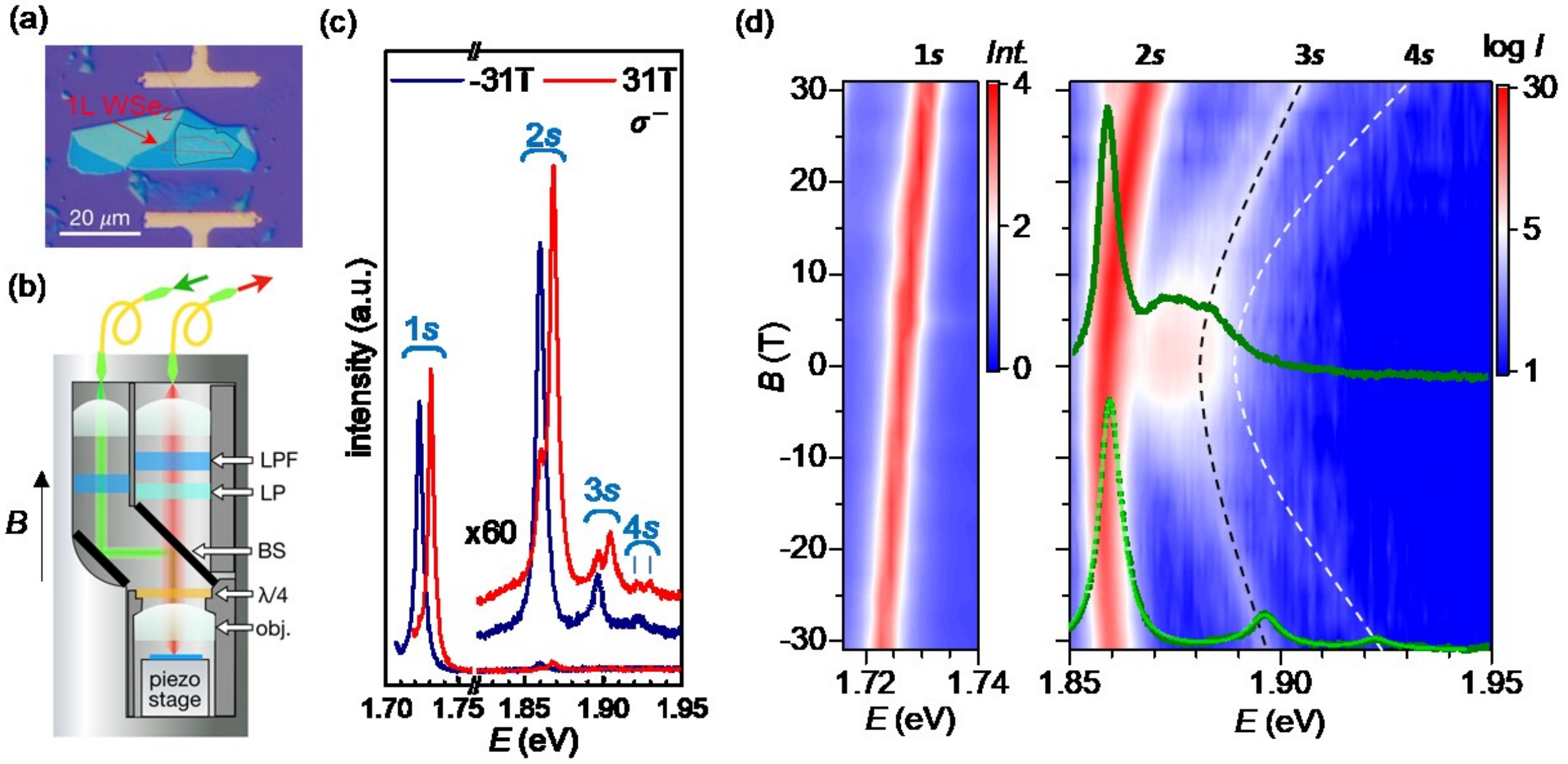

Figure 1. Magneto-PL measurement of 1*s*–4*s* excitons in 1L-$WSe_2$. (a) The optical micrograph of the hBN/1L-$WSe_2$/hBN sample. The 1L-$WSe_2$ is enclosed by the red curve. (b) Helicity resolved magneto-PL measurement setup. (c) PL spectra at ±31 Tesla. (d) The contour plot of $\sigma^-$ PL spectra as a function of the magnetic field for 1*s*, 2*s*, 3*s* and 4*s* excitons as denoted. The dashed curves on 3*s* and 4*s* excitons are guides to the eye. Overlapping are PL spectra of 1L-$WSe_2$ at 0 and –31 T.

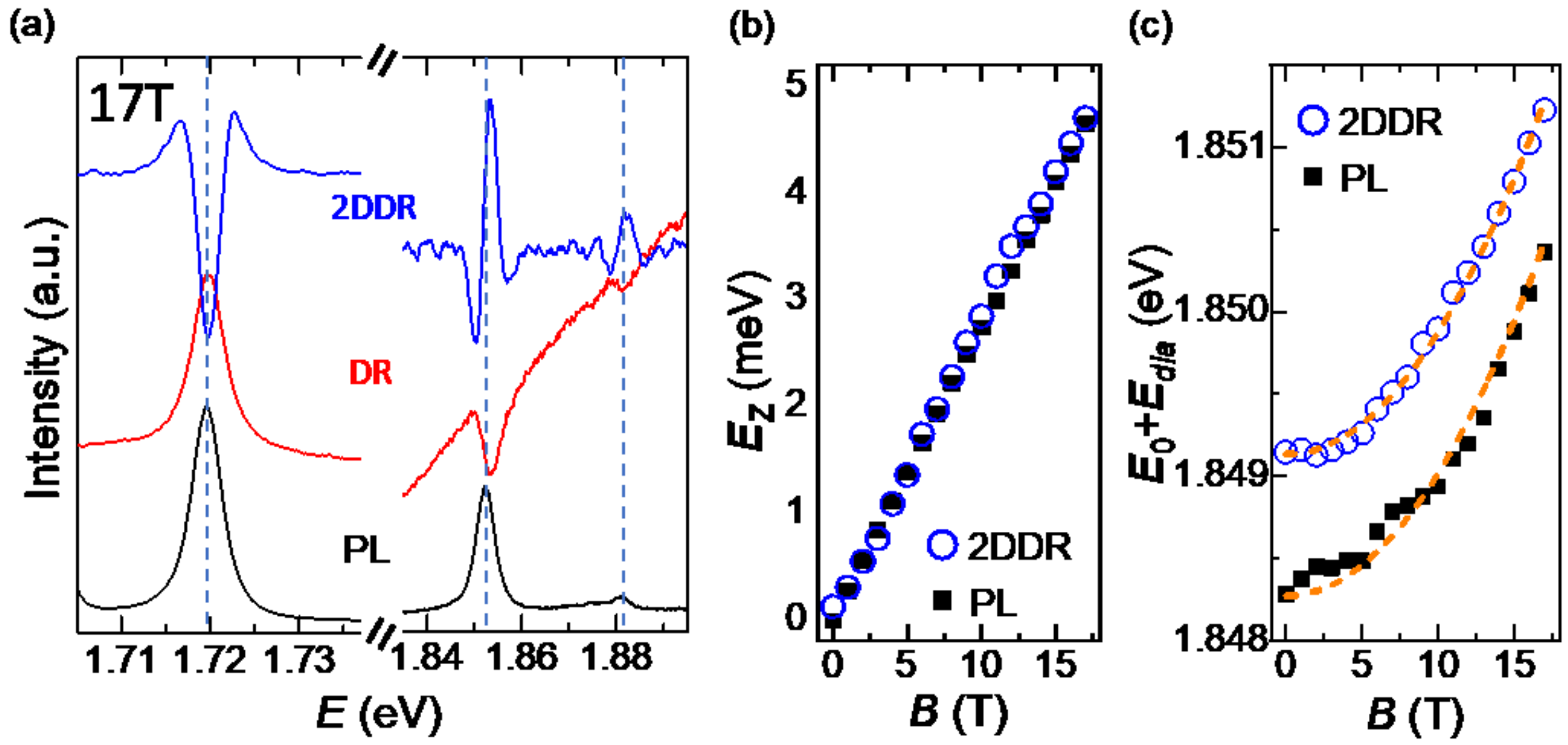

Figure 2. The consistency between PL and reflectance measurements. (a) The comparison of the PL, differential reflectance (DR) and the 2nd derivative differential reflectance (2DDR) spectra at 17 T. The dashed lines indicate the peak energy extracted from PL spectra. (b) The Zeeman shift of the 2*s* exciton extracted by PL and 2DDR spectra. (c) The magnetic field dependent average energy of 2*s* exciton $\sigma^-$ and $\sigma^+$ signal extracted by PL and 2DDR spectra. The dashed lines are quadratic diamagnetic shift fits.

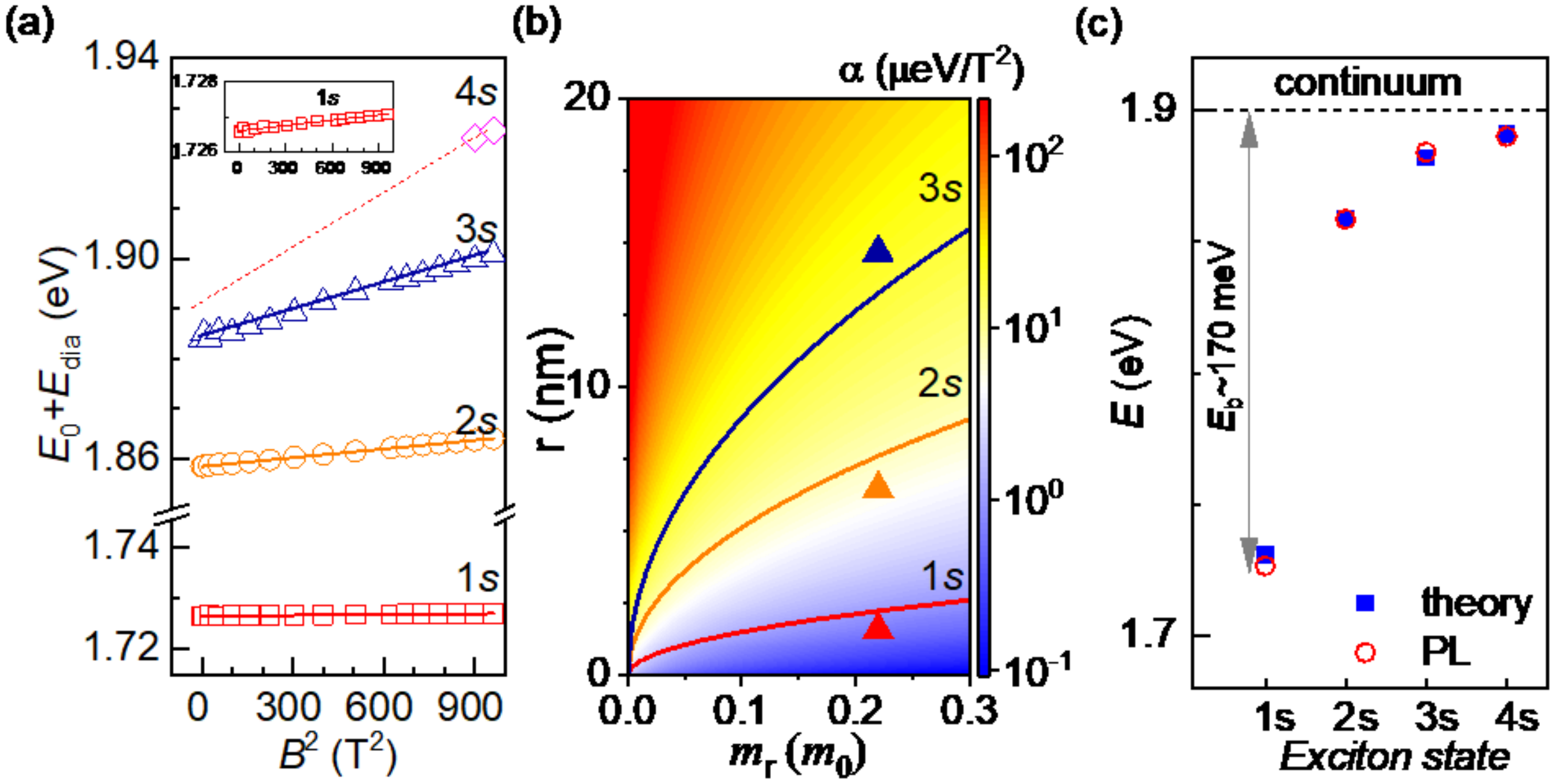

Figure 3. Rydberg exciton diamagnetic shift, size and binding energy. (a) $E_0$ + $E_{dia}$ of 1*s*–4*s* excitons plotted as a function of $B^2$. The inset shows the diamagnetic shift of 1*s* exciton in a magnified scale. (b) The radius of the exciton. The triangles are theoretical values assuming a reduced mass $m_r$ of 0.22 $m_0$. Experimentally there is not a consensus regarding $m_r$ in 1L-$WSe_2$ (see the text). In the $r$ vs. $m_r$ heatmap different colors represent different values of $\alpha = \frac{e^2}{8m_r}\langle r^2 \rangle$. The solid curves correspond to experimentally extracted $\alpha$ from the slopes in (a). (c) The Rydberg exciton energy at zero field. The black dashed line indicates the quasi-particle bandgap.

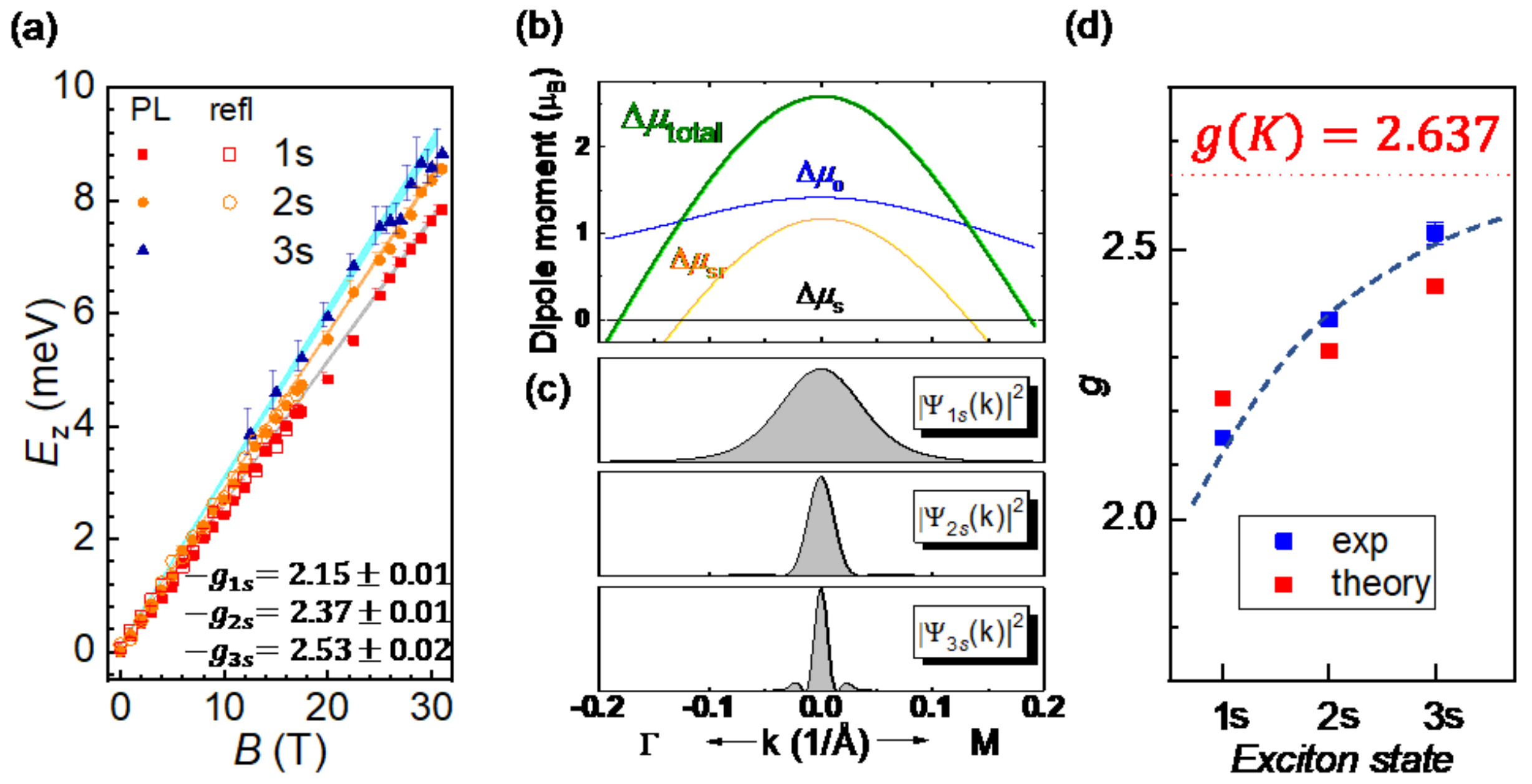

Figure 4. Evolution of Rydberg exciton magnetic dipole moment. (a) Zeeman shift of 1*s*, 2*s* and 3*s* excitons. (b) The momentum (*k*) dependence of magnetic dipole moment contributions from spin ($\Delta\mu_s$), atomic orbitals ($\Delta\mu_o$), and self-rotation ($\Delta\mu_{sr}$). (c) Spreading of the Rydberg exciton wavefunction in the momentum space. (d) Measured and calculated $g$-factors. The dotted red line indicates the value at K point, and the dashed blue curve is a guide to the eye.

# Supporting Information

## Luminescent emission of excited Rydberg excitons from monolayer $WSe_2$

*Shao-Yu Chen,[1] Zhengguang Lu,[2,3] Thomas Goldstein,[1] Jiayue Tong,[1] Andrey Chaves[4], Jens Kunstmann[5], L. S. R. Cavalcante[4], Tomasz Woźniak[5,6], Gotthard Seifert[5], D. R. Reichman,[7] Takashi Taniguchi,[8] Kenji Watanabe,[8] Dmitry Smirnov[2] and Jun Yan[1,*]*

[1]Department of Physics, University of Massachusetts, Amherst, Massachusetts 01003, USA

[2]National High Magnetic Field Laboratory, Tallahassee, Florida 32310, USA

[3]Department of Physics, Florida State University, Tallahassee, Florida, 32306, USA

[4] Departamento de Física, Universidade Federal do Ceará, Caixa Postal 6030, Campus do Pici, 60455-900 Fortaleza, Ceará, Brazil

[5] Theoretical Chemistry, Department of Chemistry and Food Chemistry & Center for Advancing Electronics Dresden, TU Dresden, 01062 Dresden, Germany

[6] Department of Theoretical Physics, Wrocław University of Science and Technology, wyb. Wyspiańskiego 27, 50-370, Wrocław, Poland

[7] Department of Chemistry, Columbia University, New York, New York 10027, USA

[8] National Institute of Materials Science, 1-1 Namiki, Tsukuba, Ibaraki 305-0044, Japan

[*]Corresponding Author: Jun Yan. Tel: (413)545-0853 Fax: (413)545-1691

E-mail: yan@physics.umass.edu

## 1. Details of theoretical modelling

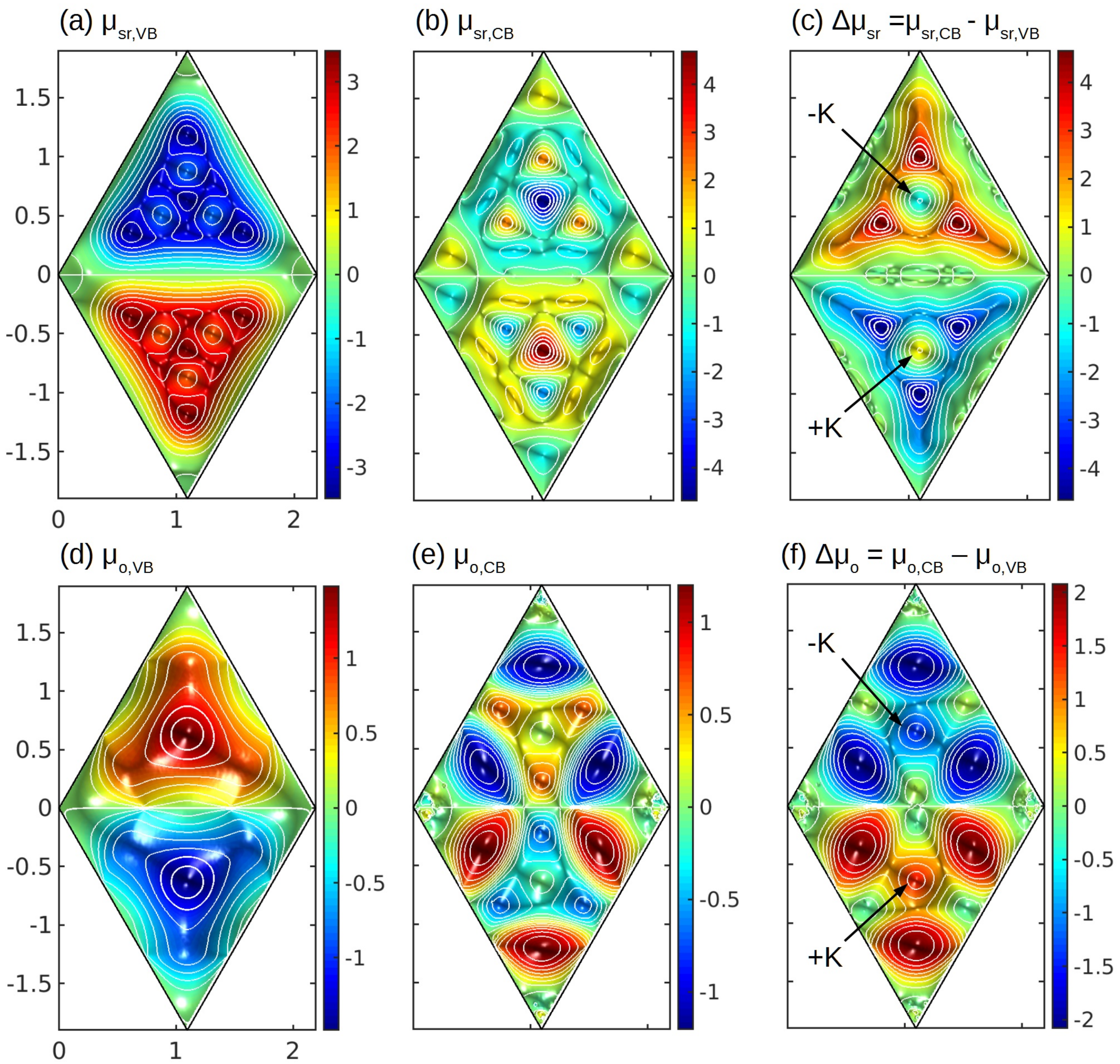

Figure S1*:* Momentum-resolved orbital magnetic moments for valence band (VB) and conduction band (CB) in $WSe_2$. First row: self-rotation contribution of (a) valence and (b) conduction band and (c) their difference. Second row: atomic orbital moments of (d) valence band and (e) conduction band and (f) their difference. The magnetic moments are given in units of $\mu_B$; the coordinates are given in units of 1/Å; the positions of the +K and -K valleys are indicated by arrows.

*Calculation of the self-rotation contribution to the magnetic moment.* According to Chang et al. [1] the angular momentum due to the self-rotation of a Bloch wave packet around its center of mass is given by

$$L^z_{n\boldsymbol{k}} = i\frac{m}{\hbar}\sum_{j\neq n}\left[\frac{\left\langle n\boldsymbol{k}\left|\frac{\partial \hat{H}}{\partial k_x}\right|j\boldsymbol{k}\right\rangle\left\langle j\boldsymbol{k}\left|\frac{\partial \hat{H}}{\partial k_y}\right|n\boldsymbol{k}\right\rangle}{E_{j\boldsymbol{k}}-E_{n\boldsymbol{k}}} - c.c.\right],$$

where $L^z_{n\boldsymbol{k}}$ is linked to the magnetic dipole moment by $\mu_{sr,n\boldsymbol{k}} = \frac{L^z_{n\boldsymbol{k}}}{\hbar}\mu_B$, $|n\boldsymbol{k}\rangle$ and $E_{n\boldsymbol{k}}$ are the lattice periodic part of the Bloch wave function and the energy of band n at the position **k** in momentum space, and $\hat{H}$, m, i are the Hamiltonian, the free electron mass and the imaginary unit, respectively. The electronic structure of $WSe_2$ was approximated by the third-nearest-neighbor, three-band tight-binding model of Liu et al. [2]. The final results, including the g-factors of 1s to 4s excitons, were determined using a 99x99x1 k-grid. The momentum-resolved moments are shown in Figure S1(a)-(c).

*Density Functional Theory Calculations of the atomic orbital contribution to the magnetic moment.* Since the 3-band model of Liu et al. is based on tungsten d-states only, the contributions from other atomic orbitals (e.g. sulfur p-states) to the electronic structure are not properly represented. We therefore used more accurate DFT to calculate the momentum-resolved atomic orbital magnetic moment throughout the Brillouine zone. This was done by projecting the Bloch states $|n\boldsymbol{k}\rangle$ of valence and conduction bands onto atom-centered real spherical harmonics $\langle alm|$, where *a* is the atom and *l, m* are the orbital and magnetic quantum numbers, respectively. The complex expansion coefficients $\langle alm|n\boldsymbol{k}\rangle$, that can be seen as an LCAO expansion of the Bloch states, were used to evaluate the expectation value of the z-component of the angular momentum operator $L^z_{n\boldsymbol{k}} = \langle n\boldsymbol{k}|\hat{L}_z|n\boldsymbol{k}\rangle = \sum_{a,l,m,m'}\left\langle n\boldsymbol{k}\middle|alm'\right\rangle\left\langle alm'\middle|\hat{L}_z\middle|alm\right\rangle\langle alm|n\boldsymbol{k}\rangle$ and the momentum matrix elements in the basis of real spherical harmonics are

$$\left\langle alm'\middle|\hat{L}_z\middle|alm\right\rangle = i\hbar\cdot\begin{bmatrix} 0 & 0 & 0 & 0 & 0 & 0 & 0 & 0 & 0 \\ 0 & 0 & 0 & 1 & 0 & 0 & 0 & 0 & 0 \\ 0 & 0 & 0 & 0 & 0 & 0 & 0 & 0 & 0 \\ 0 & -1 & 0 & 0 & 0 & 0 & 0 & 0 & 0 \\ 0 & 0 & 0 & 0 & 0 & 0 & 0 & 0 & -2 \\ 0 & 0 & 0 & 0 & 0 & 0 & 0 & 1 & 0 \\ 0 & 0 & 0 & 0 & 0 & 0 & 0 & 0 & 0 \\ 0 & 0 & 0 & 0 & 0 & -1 & 0 & 0 & 0 \\ 0 & 0 & 0 & 0 & 2 & 0 & 0 & 0 & 0 \end{bmatrix}.$$

Again $L^z_{n\boldsymbol{k}}$ is linked to the magnetic dipole moment by $\mu_{o,n\boldsymbol{k}} = \frac{L^z_{n\boldsymbol{k}}}{\hbar}\mu_B$. The relation between the magnetic quantum number and the orbitals is: m = 1(s), 2 ($p_y$), 3 ($p_z$), 4 ($p_x$), 5 ($d_{xy}$), 6 ($d_{yz}$), 7 ($d_{z2}$), 8 ($d_{xz}$), 9 ($d_{x2-y2}$). As the projection onto non-overlapping atomic spheres is incomplete (because of the voids in the interstitial space), the expansion coefficients were rescaled to obtain a normalized basis. The results are shown in Figure S1(d)-(f).

The density functional theory (DFT) calculations of $WSe_2$ monolayers were using the Perdew–Burke–Ernzerhof (PBE) functional [3]. We employed the projector augmented wave method [4] and a plane-wave basis set with a cutoff energy of 223 eV, as implemented in VASP (Vienna Ab initio Simulation Package, version 5.4.4) [5,6], using a standard PAW potential (prior to version 5.2). The k-space integration was carried out with a Gaussian smearing method using an energy width of 0.05 eV. The self-consistent-charge was calculated on a Γ-point-centered regular 6x6x1 k-grid. Unit cells of monolayers and bilayers were built with 11 Å separation between replicates in the perpendicular direction to achieve negligible interaction. All systems were fully structurally optimized. We obtained an in-plane lattice constant of 3.31 Å and a band gap of 1.34 eV (including spin-orbit interactions). The expansion coefficients $\langle alm|n\boldsymbol{k}\rangle$ were determined in a non-self-consistent-charge calculation on a 99x99x1 k-grid. For the illustrations in Fig. S1 the results were interpolated to a finer grid of 300x300x1 k-points.

We checked that spin-orbit interactions had a negligible impact on the atomic orbital moments. Therefore we did not include these effects in the final calculations.

## 2. Sample to sample variation of PL spectra

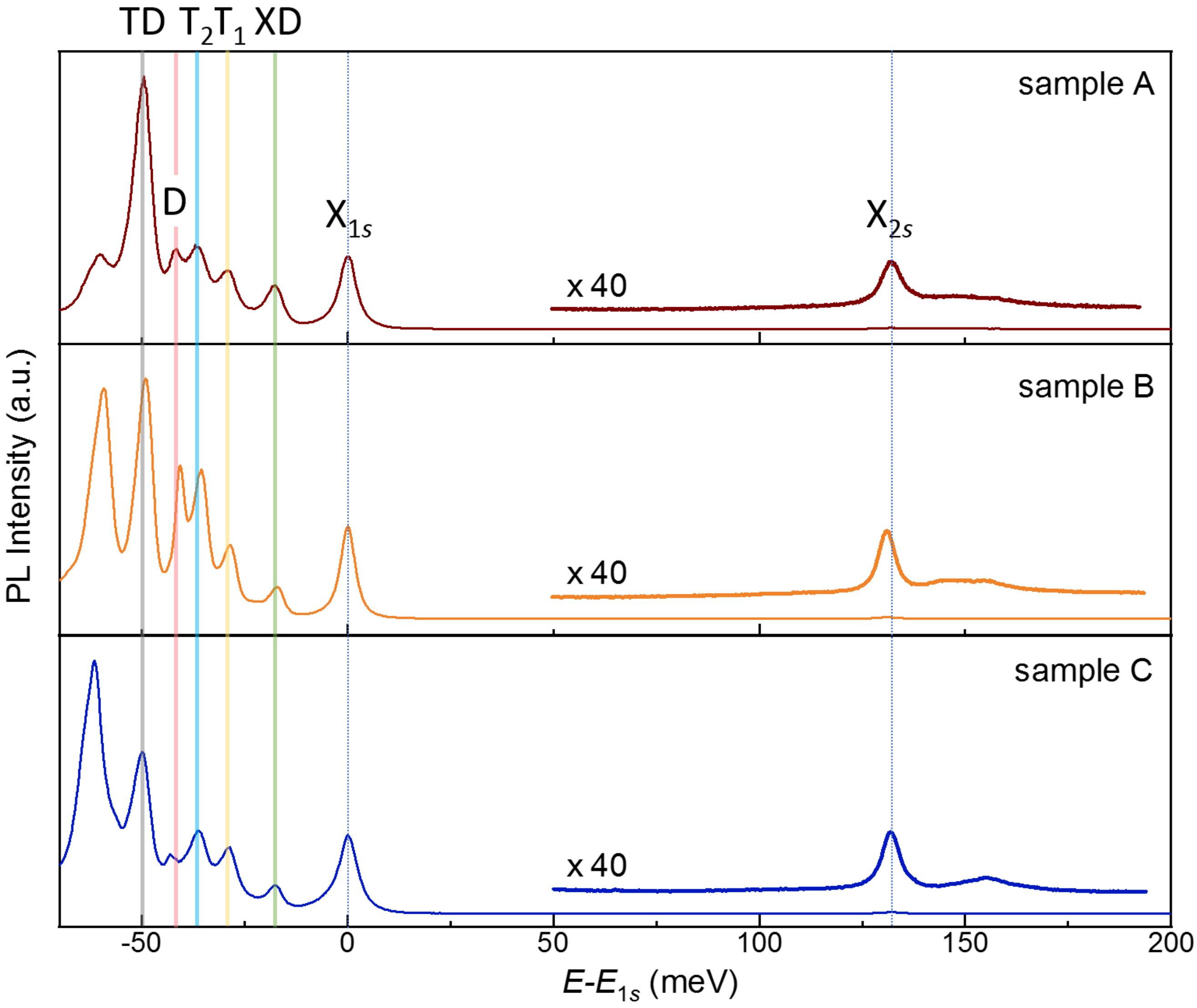

Figure S2. PL spectra of three different hBN/1L-$WSe_2$/hBN samples at 4K plotted as a function of the energy difference from 1*s* exciton ($X_{1s}$). The vertical lines are aligned to the peak energies extracted from sample A.

In this section, we present the PL spectra taken from three different samples. The 1L-$WSe_2$ is exfoliated from different flakes in the same batch of CVT growth. After exfoliation, the 1L-$WSe_2$ flake is encapsulated by two hBN flakes (~10–15 nm in thickness) via a dry transfer technique. The sandwiched samples are further thermally annealed at 350 ℃ in argon environment for 1 hour to improve the sample quality. Note that trapped air bubbles and polymer residue can cause nonuniformity and spot-to-spot variation. In Figure S2, we plot the PL spectra of the best spots of the three samples as a function of energy difference from 1*s* exciton ($X_{1s}$). The absolute energy and linewidth of $X_{1s}$ in these samples exhibit slight variations: 1.727 eV (5.1 meV), 1.717 eV (4.3 meV) and 1.736 eV (5.9 meV) for sample

A, B and C, respectively. In the lower energy, we consistently observe several intrinsic peaks from multi-particle bound exciton states. Here, we mark the peaks in the convention in our recently published paper.[7] From the right to the left, there are emission from biexciton (XD), two negatively charged trions ($T_1$ and $T_2$), the dark exciton (D) and the five-particle exciton-trion (TD). The appearance of $T_1$ and $T_2$ as well as TD indicates that the sample is slightly electron-doped. At energies higher than $X_{1s}$, all samples display clear 2*s* exciton emission ($X_{2s}$) and a broad feature on the high energy side as discussed in the main text.

## Supplementary References